# J-type carbon stars: A dominant source of $^{14}$N-rich presolar SiC grains of type AB

Nan Liu[1*], Thomas Stephan[2,3], Patrick Boehnke[2,3], Larry R. Nittler[1], Conel M. O'D. Alexander[1], Jianhua Wang[1], Andrew M. Davis[2,3,4], Reto Trappitsch[2,3,5], Michael J. Pellin[2,3,4,6]

[1]Department of Terrestrial Magnetism, Carnegie Institution for Science, Washington, DC 20015, USA; nliu@carnegiescience.edu

[2]Department of the Geophysical Sciences, The University of Chicago, Chicago, IL 60637, USA

[3]Chicago Center for Cosmochemistry, Chicago, IL, USA

[4]The Enrico Fermi Institute, The University of Chicago, Chicago, IL 60637, USA.

[5]Nuclear and Chemical Sciences Division, Lawrence Livermore National Laboratory, Livermore, CA 94550, USA.

[6]Materials Science Division, Argonne National Laboratory, Argonne, IL 60439, USA

ABSTRACT

We report Mo isotopic data of 27 new presolar SiC grains, including 12 $^{14}$N-rich AB ($^{14}N/^{15}N > 440$, AB2) and 15 mainstream (MS) grains, and their correlated Sr and Ba isotope ratios when available. Direct comparison of the data for the MS grains, which came from low-mass asymptotic giant branch (AGB) stars with large *s*-process isotope enhancements, with the AB2 grain data demonstrates that AB2 grains show near-solar isotopic compositions and lack *s*-process enhancements. The near-normal Sr, Mo, and Ba isotopic compositions of AB2 grains clearly exclude born-again AGB stars, where the intermediate neutron-capture process (*i*-process) takes place, as their stellar source. On the other hand, low-mass CO novae, and early R- and J-type carbon stars show $^{13}$C and $^{14}$N excesses but no *s*-process enhancements and are thus potential stellar sources of AB2 grains. Since both early R-type carbon stars and CO novae are rare objects, the abundant J-type carbon stars (10−15% of all carbon stars) are thus likely to be a dominant source of AB2 grains.

## 1. INTRODUCTION

Among the starting materials of the solar system were dust grains that formed in dying stars of varying types and had become incorporated in the protosolar molecular cloud. A small fraction of these presolar grains survived destructive processes in the early solar nebula and were later incorporated into the parent bodies of primitive meteorites. They are identified in laboratories on Earth by their highly anomalous isotopic compositions that resulted from progenitor stellar nucleosynthesis (e.g., Zinner 2014; Nittler & Ciesla 2016). Silicon carbide (SiC) is the most extensively studied presolar phase, and presolar SiC grains have been divided into five main groups (each of which comprises ≥1% of presolar SiC) according to their C, N, and Si isotope ratios (Hoppe et al. 1994). SiC grains of type AB (5−6% of SiC) are defined as grains with large $^{13}$C enrichments ($^{12}C/^{13}C<\sim10$) and are the second most common group after mainstream (MS) SiC grains (>90% of SiC) from low-mass asymptotic giant branch (AGB) stars (e.g., Lugaro et al. 2003; Liu et al. 2015).

Of the various types of presolar SiC, AB grains, however, have the most ambiguous stellar origin(s) and several types of stars have been proposed as their progenitors. The AB grains were initially divided into two groups, A and B (Hoppe et al. 1994; $^{12}C/^{13}C_A<3.5\leq^{12}C/^{13}C_B\leq10$), which were subsequently combined as a single group with $^{12}C/^{13}C\leq10$ due to the lack of any other obvious differences (Amari et al. 2001). A recent study of AB grains (Liu et al. 2017) has shown distinctive differences in the isotopic compositions of C, N, Al, and Si between $^{15}N$-rich ($^{14}N/^{15}N \leq 440$, the solar ratio, AB1 group) and $^{14}N$-rich ($^{14}N/^{15}N > 440$, AB2 group) AB grains, thus suggesting different stellar sources of these two subgroups. Liu et al. (2017) found that AB1 grains (2−3% of all SiC) show an anticorrelation between their $^{14}N/^{15}N$ and their inferred $^{26}Al/^{27}Al$ ratios, which points to an origin in supernovae, in which H was ingested into the He/C zones prior to the explosions. In comparison, AB2 grains (2−3% of all SiC) do not show any correlation between their N and Al isotope ratios, and their stellar origins remain unclear. The previously proposed potential stellar sources for AB2 grains include born-again AGB stars, J-type carbon stars, and low-mass CO novae (Amari et al. 2001; Fujiya et al. 2013; Haenecour et al. 2016). Born-again AGB stars are post-AGB stars that experienced a late thermal pulse (TP) and thus show stronger or at least similar *s*-process isotopic signatures compared to AGB stars, according to both observations and model calculations (e.g., Busso et al. 2001; Herwig et al. 2011). In contrast, observations of J-type carbon stars show almost no detectable *s*-process enhancements (Abia & Isern 2000), and *s*-process nucleosynthesis does not occur during nova explosions (e.g., José 2016). Isotopic compositions of *s*-process elements in AB2 grains are therefore key for probing the stellar source(s) of AB2 grains.

Information on the abundances and isotopic compositions of *s*-process elements in AB2 grains, however, is extremely limited. Although Amari et al. (1995, 2001) reported abundances of *s*-process elements in 21 AB grains, only one of the grains they analyzed was an AB2 grain, and the other 20 grains were either AB1 grains or remain ambiguous, because their $^{14}N/^{15}N$ ratios of 450−480 lay close to the solar value. Later, Savina et al. (2003) measured Mo isotope ratios in seven AB grains and found normal Mo isotopic composition in five of them. However, only two of their grains belonged to the AB2 group, with one of them showing slight, ~150‰, *p*-process enrichments, i.e., $^{92,94}Mo$ excesses. Isotopic compositions of *s*-process elements in three additional AB2 grains were reported in the literature (Pellin et al. 2000; Savina et al. 2003; Barzyk et al. 2008), which had either normal Mo and Ba isotopic compositions or small depletions in $^{96}Zr$. However, all the isotopic data were only briefly reported in the abstracts, which did not discuss the possibility of surface contamination causing the normal isotopic compositions. As a result, it remains unclear if AB2 grains indeed lack *s*-process isotopic signatures.

To better understand the stellar origin(s) of AB2 grains, we report Mo isotopic compositions of 27 presolar SiC grains, including 12 AB2 and 15 MS grains, and their correlated Sr and Ba isotopic compositions when available. Our analysis of the AB2 and MS grains on the same sample mounts in the same analytical session allowed direct comparison of their *s*-process isotopic compositions, which demonstrates that AB2 grains show weaker *s*-process to solar isotopic signatures relative to MS grains from low-mass AGB stars with near-solar metallicities.

## 2. METHODS

The SiC grains in this study were extracted from the primitive Murchison (CM2) meteorite using the chemical method described in Nittler & Alexander (2003). Isolated and size-separated (~1 μm) grains in a water suspension were dispersed onto three gold mounts labeled #1, #2, and #3. SiC grains on the three mounts were identified by automated BSE (backscattered electron)-EDX (energy-dispersive X-ray) analysis (Liu et al. 2016a). Isotopic analysis of C, N, and Si in the AB2 and MS grains was performed with the NanoSIMS 50L ion microprobe at the Carnegie Institution using standard procedures (Liu et al. 2016b, 2017). All data were collected in the imaging mode, with ~150 nm spatial resolution, which allowed selection of smaller regions of interest during data reduction to minimize potential contamination. After the NanoSIMS analyses, a focused $Ga^+$ ion beam (FIB) was used to remove potential contaminants in ≥3×3 $\mu m^2$ areas adjacent to the AB2 grains at low beam current (~50 pA).

The 12 FIB-cleaned AB2 and 15 well-isolated MS grains (not FIB-cleaned) were then analyzed with a new resonance ionization mass spectrometry (RIMS) instrument, the Chicago Instrument for Laser Ionization (CHILI), at the University of Chicago. CHILI is equipped with six Ti:sapphire lasers for resonance ionization (Stephan et al. 2016). All of the Sr, Mo, and Ba isotopes were simultaneously measured in each grain using the procedure described by Stephan et al. (2017). A 351 nm Nd:YLF laser beam, focused to ~1 μm, was used to desorb materials from the surface of presolar grain mounts, so the FIB-cleaning ensured minimal sampling of potential contamination from the surrounding areas.

The NanoSIMS and CHILI data are reported in Table 1; uncertainties in Table 1 and in all figures are 2σ.[1] The C and N isotope data are reported as atomic ratios, and the Si, Sr, Mo, and Ba isotope data as δ-values[2]. The Si, Sr, Mo, and Ba isotope data were normalized to $^{28}$Si, $^{87}$Sr, $^{96}$Mo, and $^{136}$Ba, respectively. Although $^{86}$Sr is commonly used for normalizing Sr isotope ratios in the literature (e.g., Liu et al. 2015), we chose to normalize all the Sr isotope ratios to $^{87}$Sr, because an unidentified molecular interference peak appeared at mass 86 in some of the RIMS spectra of grains. Values of $\delta^{130,132}$Ba and $\delta^{84}$Sr (in most cases) are not reported in Table 1 because of large statistical uncertainties caused by the low abundances of these isotopes. Strontium and Ba isotope ratios were obtained in 15 and 10 grains, respectively, out of 27 grains with available Mo isotope ratios (Table 1), indicating lower efficiencies of the Sr and Ba resonance ionization schemes and/or lower Sr and Ba concentrations in the grains.

## 3. RESULTS

Figure 1 shows that the MS grain data from this study are in good agreement with the literature data[3] and the data for N stars, consistent with the fact that MS grains came from low-mass AGB stars (C (N)-type carbon stars) with near-solar metallicities, and are thus characterized by *s*-process isotopic signatures (e.g., Lugaro et al. 2003; Liu et al. 2015).

[1]Note that 3 of the 12 AB2 grains had $^{12}$C/$^{13}$C ratios of 11, which are slightly higher than the definition of AB2 grains, $^{12}$C/$^{13}$C ≤10. However, as will be shown later, these three AB2 grains had near-normal Mo isotopic compositions as the other nine AB2 grains, but different from the anomalous Mo isotopic compositions observed in most of the MS grains, which thus supports the relaxation of the AB grain definition as also suggested by Liu et al. (2017).

[2] δ-notation is defined as $\delta^{i}A=[(^{i}A/^{j}A)_{grain}/(^{i}A/^{j}A)_{std}-1]\times 1000$, where A denotes an element, i an isotope of this element, and j the normalization isotope, and $(^{i}A/^{j}A)_{grain}$ and $(^{i}A/^{j}A)_{std}$ represent the corresponding isotope ratios measured in a sample and the standard, respectively.

[3]Note that the different MS grain datasets in the literature (e.g., Hynes & Gyngard 2009) show varying $^{14}$N/$^{15}$N distributions, indicating different degrees of solar/terrestrial N contamination sampled during ion probe analyses. Thus, a large dataset from Gyngard et al. (2006) that shows a relatively low degree of N contamination was chosen for comparison in Fig. 1.

Intriguingly, Fig. 1 shows that AB2 grains with $^{14}N/^{15}N \geq 1{,}000$ (hereafter high-$^{14}N$ AB2 grains) in this study span a narrower $^{12}C/^{13}C$ range than those with $^{14}N/^{15}N<1{,}000$ (hereafter low-$^{14}N$ AB2 grains). This observation seems to be supported by the literature AB2 grain data[4]. It is unclear why the high-$^{14}N$ AB2 grains from this study lie at the high $^{12}C/^{13}C$ end; although the higher $^{12}C/^{13}C$ ratios obtained in our study could indicate a higher degree of C contamination, the consistent $^{12}C/^{13}C$ range for the low-$^{14}N$ AB2 grains from this study and the literature seems not to favor this possibility. On the other hand, the AB2 and MS grains in this study show a similar range of $\delta^{29}Si$ values indicating a similar range of progenitor stellar compositions, with the MS grains being more $^{30}Si$-enriched.

The Mo isotopic data of 12 AB2 and 15 MS grains are shown in Fig. 2, in comparison to AGB model predictions for a 2 $M_\odot$, 0.5 $Z_\odot$ star (Gallino et al. 1998; Liu et al. 2015; Bisterzo et al. 2017). The AB2 grains clearly show near-solar Mo isotopic compositions. In contrast, highly anomalous Mo isotope ratios were obtained in 12 out of the 15 MS grains and can be well explained by the range of neutron exposures constrained by the correlated Sr and Ba isotope ratios of MS grains reported by Liu et al. (2015, Table 6). In addition, the most anomalous Mo isotope pattern observed in the 12 AB2 grains is similar to the averaged *s*-process pattern of the 12 MS grains that had anomalous Mo isotope ratios within 2σ uncertainties, but seems to lie closer to the solar Mo isotopic composition (Fig. 3). Note that 16 AB1 grains were also measured in the same CHILI session for Sr, Mo and Ba isotopes. Although the data will be reported elsewhere, it is worth mentioning that eight out of the 16 AB1 grains show *s*-process Mo isotope ratios with slightly more anomalous values, thus providing further support to the division of the AB grains into two subgroups.

## 4. DISCUSSION

### *4.1. Can Contamination Explain AB2 Mo Isotope Data?*

Multielement isotope analyses of heavy elements in Murchison presolar SiC grains by Barzyk et al. (2007) indicated substantial contamination of Ba and Mo. Thus, an acid-cleaning method was used in subsequent studies of Sr and Ba isotopes in SiC (Liu et al. 2014b, 2015); this method was demonstrated to be quite effective in removing surface Sr and Ba contamination based on comparison with the literature data. Since the Murchison SiC grains in the present study were not further acid-cleaned using this method, potential surface contamination caused by parent-body processes and/or laboratory contamination could result in dilution of anomalous isotopic signatures measured in grains from this study. Liu et al. (2014b, 2015) also pointed out that MS grains with $\delta^{135}Ba$ values above −400‰ are likely to be Ba-contaminated grains.

Thus, we tested the cleanliness of our MS grains by comparing (Fig. 4) the grain data in this study to the same set of AGB model predictions in Fig. 2 using the method introduced by Barzyk et al. (2007). In the Barzyk et al. study, a "contaminated" grain is defined as one that cannot be reached by various AGB model predictions in the C-rich phase within 2σ error ellipses in a two-element isotope plot. Based on this criterion, at least one of the three grains with normal Mo isotope ratios is Mo-contaminated. Since "clean" MS grains are likely to have $\delta^{135}Ba$ values below −400‰, most likely all three grains are Mo-contaminated, thus corresponding to 20% of the grains having severe Mo contamination. Note that we also infer similar levels of Mo contamination from correlated Mo and Ba data of 21 Y (14%) and 21 Z (14%) grains from lower-metallicity AGB stars analyzed in the same CHILI session; data for these grains will be reported elsewhere. It is also worth pointing out that the rare-type grains including AB1, AB2, Y, and Z grains analyzed in this study were all FIB-cleaned, but the MS grains were not. Thus, the MS grain data should represent the highest level of surface contamination in this study. In addition, only one out of 16 MS, Y, and Z grains had $\delta^{135}$Ba values higher than −400‰, also indicating a negligible amount of Ba contamination in our study. Finally, the level of Sr contamination in this study is unclear, because the set of AGB models in Fig. 2 predict $\delta^{88}$Sr (normalized to $^{87}$Sr) values to lie close to the solar value with small variations (ranging from −40‰ to 300‰ in D6 to D2 cases, respectively). Thus, the normal $\delta^{88}$Sr values measured in 19 MS, Y, and Z grains could be consistent with either *s*-process nucleosynthesis and/or solar Sr contamination. Although negative $\delta^{84}$Sr values are diagnostic of *s*-process nucleosynthesis (a pure-*s* component would have $\delta^{84}$Sr=−1,000‰), $\delta^{84}$Sr values were obtained in only a few grains due to the low $^{84}$Sr abundance. To summarize, multielement isotopic analyses of a number of MS, Y, and Z grains show that only up to 20% of the grains suffered from severe Mo contamination, and that Ba contamination seems to be negligible.

[4] The AB2 literature data include all AB2 grains reported in the literature except for those from Amari et al. (2001), whose AB2 grains do not show a narrower range of $^{12}C/^{13}C$ ratios for $^{14}N/^{15}N>1{,}000$ as observed in other datasets.

In comparison, 10 out of 12 AB2 grains had normal Mo isotope ratios, and the other two grains had much weaker *s*-process Mo isotopic signatures with large uncertainties relative to MS grains (Figs. 2,3). Although AB2 grains seem to show lower concentrations of *s*-process elements, the Mo counts collected from the 10 AB2 grains with normal isotopic compositions (median: 3,500 counts) lie close to the ranges of MS (5,400 counts), Y (5,600 counts), and Z (5,000 counts) grains. Therefore, the level of Mo contamination should be similar for the different types of SiC grains measured in this study, which is strongly supported by the similar levels of Mo contamination inferred from the MS, Y, and Z grains. By considering the highest level of Mo contamination inferred from the MS grain data (20%), the normal Mo isotope ratios in only up to 2 or 3 out of the 12 AB2 grains could be explained by Mo contamination.

More importantly, AB2 and MS grains barely overlap in Fig. 2, as the former are mainly clustered around the solar composition, while the latter lie closer to the *s*-process end-member in this Mo three-isotope plot. Specifically, the averaged $\delta^{92}$Mo value for the 12 AB and 15 MS grains are −42±50‰ and −687±23‰, respectively, demonstrating that the two groups of grains are distinctively different in their Mo isotopic compositions. Finally, except for grain M2-A1-G554, which had the most anomalous Mo isotope ratios (Fig. 3) and $\delta^{88}$Sr=−817±226‰, all five other AB2 grains had solar Sr and Ba isotope ratios (Table 1), providing further support to the indication that the AB2 grains had near-normal isotopic compositions for *s*-process elements. The extremely low $\delta^{88}$Sr value observed in M2-A1-G554 likely indicates that the branch points at $^{85}$K and $^{86}$Rb were not activated at all, so that the *s*-process likely proceeded following the path $^{85}$Kr–$^{85}$Rb–$^{86}$Rb–$^{86}$Sr, resulting in the low $\delta^{88}$Sr value (path 1 in Fig. 2 of Liu et al. 2015). Since the lowest $\delta^{88}$Sr value that can be reached by AGB models is around −300‰ by turning off the $^{22}$Ne source (e.g., Lugaro et al. 2003), the $\delta^{88}$Sr value of this grain suggests a neutron-capture environment with an extremely low peak neutron density in its progenitor star relative to the typical *s*-process in AGB stars.

### *4.2. Potential Stellar Origin(s) of AB2 Grains*

The near-normal isotopic compositions of AB2 grains clearly exclude born-again AGB stars as their stellar sources. Dust grains from born-again AGB stars are expected to show neutron-capture isotopic signatures as a result of the intermediate neutron-capture process (*i*-process) at enhanced neutron densities (~$10^{15}$ cm$^{-3}$, Herwig et al. 2011) relative to those of the *s*-process (~$10^{7}$ cm$^{-3}$, Gallino et al. 1998). As a result, born-again AGB stars have been proposed as the stellar source of $^{13}$C-rich graphite grains with large $^{42,43,44}$Ca excesses (Jadhav et al. 2013). The nucleosynthetic model of Herwig et al. (2011) for the born-again AGB star, Sakurai's object, predicts huge excesses of $^{87,88}$Sr, $^{96,97,98,100}$Mo, and $^{136}$Ba, and thus extremely anomalous Sr, Mo, and Ba isotopic compositions as a result of the *i*-process nucleosynthesis (F. Herwig & M. Pignatari, 2014, private communication)[5]. In contrast, the AB2 grains in this study show neutron-capture signatures that are even weaker than the *s*-process signatures and are thus inconsistent with being from born-again AGB stars (Liu et al. 2014b). It should be pointed out that none of the AB1 grains have been clearly demonstrated to have come from born-again AGB stars (Liu et al. 2016b, 2017). On the other hand, the graphite grains that have been assigned to a born-again AGB stellar origin (Jadhav et al. 2013) could be alternatively explained by nucleosynthesis in supernovae with H ingestion (Liu et al. 2017). One of the most diagnostic isotope ratios for the two stellar sources, $^{14}$N/$^{15}$N (Liu et al. 2017), however, is terrestrial in most of the graphite grains, probably due to surface contamination and/or isotope equilibration in the Orgueil meteorite parent body (Jadhav et al. 2013). Thus, the contribution of born-again AGB stars to the dust reservoir in the early solar system remains unclear. However, it is noteworthy that although more than 10% of low- and intermediate-mass stars that go through a planetary nebula phase will become born-again AGB stars based on theoretical estimates (Iben et al. 1996), the observed amount of C-rich dust produced by Sakurai's object ($6\times10^{-5}$ $M_{\odot}$, Chesneau, et al. 2009) is generally substantially lower than the amounts of C-rich dust ($10^{-4}$–$10^{-2}$ $M_{\odot}$) produced by low-mass AGB carbon stars (e.g., Zhukovska & Henning, 2013).

Low-mass CO novae and early R- and J-type carbon stars are all potential stellar sources of AB2 grains because they are enriched in $^{13}$C and $^{14}$N, and lack *s*-process enhancements according to astronomical observations and/or calculations (Dominy 1984; Abia & Isern 2000; Zamoro et al. 2009; Jordi et al. 2012; Hedrosa et al. 2013). Both early R- and J-type carbon stars have near-solar metallicities, with the former being sparse in number (Zamora et al. 2009) and the latter being abundant with higher $^{13}$C and Li enrichments (10−15% of all carbon stars, Abia & Isern 2000). The observed chemical properties of early R- and J-type carbon stars are consistent with those of AB2 grains, including their C and N isotope ratios, their lack of *s*-process enhancements, and their near-solar metallicities (Figs. 1,2). Also, J-stars are quite abundant amongst carbon stars and, therefore, should have contributed a large amount of C-rich dust (more than 10% by taking the relative ratio of J-stars to N-stars) to the early solar system. Thus, all these facts strongly indicate that a majority of AB2 grains came from J-type carbon stars.

Nova nucleosynthesis in the context of presolar grain data has been described in detail by José et al. (2004) and José & Hernanz (2007). Overall, the 0.6 $M_{\odot}$ CO nova model predictions are in good agreement with the AB2 grain data. First, the model predicts that Si isotope ratios are unaffected by the nova nucleosynthesis during the nova explosion (José et al. 2004), which is consistent with the fact that AB2 grains are less enriched in $^{30}$Si relative to MS grains, because the *s*-process nucleosynthesis in AGB stars that enriches $^{30}$Si in MS grains does not occur in low-mass CO novae (Fig. 1); thus, the Si isotope ratios of AB2 grains best represent the Galactic chemical evolution trend of Si isotope ratios (Fig. 1). Second, the predicted C and Al isotope ratios in the ejecta of 0.6 $M_{\odot}$ CO novae (Table 2 of José et al. 2004) are slightly more extreme than the AB2 grain data (Fig. 1; Amari et al. 2001), but this problem can be reconciled by invoking mixing with material with close-to-normal isotopic composition from the companion main-sequence star. Finally, the model predicts $^{14}$N/$^{15}$N ratios only up to 2,000, which cannot cover the whole $^{14}$N/$^{15}$N range of AB2 grains (up to ~10,000), and dilution with material from the companion star is likely to lower the $^{14}$N/$^{15}$N ratios. So, low-mass CO novae alone do not seem to be able to explain all the AB2 grain data unless there exist lower-than 0.6 $M_{\odot}$ CO novae capable of producing $^{14}$N/$^{15}$N ratios of up to 10,000.

---

[5] Note that although Sakurai's object has a subsolar Ba elemental abundance, the Herwig et al. model predicts strong anomalous Ba isotopic compositions as a result of the *i*-process (Liu et al. 2014b).

Additionally, novae and early R-type stars are extremely rare, while, in contrast, J-type carbon stars are quite common objects. Calculations have shown that novae contributed only 0.1% to the dust inventory in the interstellar medium, while the contribution of carbon stars is 200 times higher (Gehrz 1989). Thus, it is very unlikely that low-mass CO novae are a dominant source of AB2 grains, which make up 2−3% of presolar SiC grains. Interestingly, only low-$^{14}$N AB2 grains show $^{12}$C/$^{13}$C ratios lower than five, likely indicating an additional stellar source for low-$^{14}$N AB2 grains that resulted in the wider range of $^{12}$C/$^{13}$C ratios compared to the high-$^{14}$N AB2 grains. Based on this observation and the occurrence frequency of CO novae in the Galaxy, it is likely that low-mass CO novae only contributed a very small fraction of AB2 grains to the early solar system (Haenecour et al. 2016).

Traditionally, J-type carbon stars are proposed to be the *daughters* of early-R stars (e.g., Lloyd Evans 1986) because of their chemical similarities and the higher luminosity of J stars. However, the two types of carbon stars have different Galactic distributions, which argues against a genetic relationship between the two (Zamora et al. 2009). As discussed by Zamora et al. (2009) in detail and investigated in many recent studies (Izzard et al. 2007; Zhang & Jeffery 2013; Sengupta et al. 2013), the most likely scenario to explain the optical and chemical properties of early R- and J-type stars are binary mergers with the most favorable cases being He white dwarf and red giant mergers (Izzard et al. 2007; Zhang & Jeffery 2013). However, so far, the investigated merger channels are not able to account for the observed populations (in both absolute and relative abundances) of early R- and J-type carbon stars (Zhang & Jeffery 2013; Sengupta et al. 2013), and further simulation efforts are needed to investigate other merger channels along with more detailed nucleosynthetic calculations that can be constrained by the AB2 grain data.

## 5. CONCLUSIONS

We obtained Mo isotope ratios in 12 AB2 and 15 MS SiC grains with correlated Sr and Ba grain data in subsets of these grains. Direct comparison of the MS grain data with the AB2 grain data from this study demonstrates that normal Mo isotope ratios, the rule rather than the exception in AB2 grains, represent an intrinsic property of the grains set by their progenitor stars instead of laboratory Mo contamination, and this is further supported by the correlated Sr and Ba isotope data. The lack of *s*-process isotopic signatures in AB2 grains clearly points to potential stellar sources of low-mass novae, and early R- and J-type carbon stars, and of these, J-type carbon stars are the most abundant observed stellar objects with the other two being rare. Thus, a majority of AB2 grains most likely originated in J-type carbon stars with minor contributions from low-mass (≤0.6 $M_{\odot}$) novae and early R-type carbon stars.

Interestingly, we observed that for AB2 grains, $^{12}C/^{13}C$ ratios lower than five are only seen in low-$^{14}N$ AB2 grains ($^{14}N/^{15}N$<~1,000), which could have come from low-mass CO novae that are predicted to produce relatively low $^{12}C/^{13}C$ (<2) and $^{14}N/^{15}N$ (<2,000) ratios. The fact that none of the $^{13}C$-rich SiC grains seem to have originated from ONe novae is also consistent with the observation that high-mass ONe novae are usually dust-free while low-mass CO novae are dusty objects (Gehrz 1989). Finally, J-type carbon stars are still poorly understood stellar objects, so detailed isotopic signatures of AB2 grains should provide stringent tests to evolutionary simulations of J-type carbon stars that need to be considered by stellar modelers.

Acknowledgements: We thank the anonymous referee for a constructive reading of the manuscript. This work was supported by NASA through grants NNX10AI63G to L.R.N. and NNX15AF78G to A.M.D.. We thank Dr. Sara Bisterzo and Prof. Roberto Gallino for providing the Torino AGB model calculations used in this study. Part of this work was performed under the auspices of the U.S. Department of Energy by Lawrence Livermore National Laboratory under Contract DE-AC52-07NA27344. LLNL-JRNL-732767.

## FIGURE CAPTIONS

**Fig. 1.** The isotopic compositions of C, N, and Si of AB2 and MS SiC grains from this study are compared to literature data (Gyngard et al. 2006; Hynes & Gyngard 2009) and observational data for J- and N-type carbon stars (Hedrosa et al. 2013). For $^{14}N/^{15}N$, the protosolar value of 441±5 reported by Marty et al. (2011) is shown as a dash-dotted line for reference. The dashed lines represent terrestrial isotopic compositions. The literature data for AB2 and MS grains (Hynes & Gyngard 2009) were used for the Si-isotope linear fits. For clarity, the uncertainties in the literature presolar grain data and the observational data for N stars are not shown. The observational data with arrows are lower limits.

**Fig. 2.** A Mo three-isotope plot of $\delta^{94}$Mo versus $\delta^{92}$Mo for SiC AB2 and MS grains from this study and Torino AGB model predictions (Gallino et al. 1998). The entire evolution of the AGB envelope composition is shown, but symbols are plotted only for third dredge-up episodes that have raised the C/O ratio above 1, when SiC can condense according to thermodynamic equilibrium calculations (Lodders & Fegley 1995). The detailed model description is given in Liu et al. (2014a, 2015), and the D2 to D6 cases in the Zone-II_P4 model for a 2 $M_{\odot}$, 0.5 $Z_{\odot}$ AGB star are chosen for comparison, because they best match the correlated Sr-Ba MS grain data of Liu et al. (2015). Briefly, the Zone-II_P4 model adopts a flat $^{13}$C profile in the so-called $^{13}$C-pocket with four times higher mass relative to the standard one and the D2 to D6 cases correspond to a range of $^{13}$C concentrations within the pocket. Note that the model considers the effect of Galactic chemical evolution and as a result, the predictions start from positive values instead of the solar composition.

**Fig. 3.** The Mo isotope pattern of the most anomalous AB2 grain M2-A1-G554, in comparison to the averaged pattern of the 12 MS grains with anomalous Mo isotope ratios.

**Fig. 4.** Plot of $\delta^{135}$Ba versus $\delta^{92}$Mo for MS SiC grains from this study and the same set of Torino AGB model predictions in Fig. 2.

**Table 1.** Isotopic Data of AB2 and MS Grains. All data are reported with 2σ errors.

**Table 1.** Isotopic Data of AB and MS Grains. All the data are reported with 2σ errors.

| Grain | Group | Size ($\mu m^2$) | $^{12}C/^{13}C$ | $^{14}N/^{15}N$ | $\delta^{29}Si$ (‰) | $\delta^{30}Si$ (‰) | $\delta^{92}Mo$ (‰) | $\delta^{94}Mo$ (‰) | $\delta^{95}Mo$ (‰) | $\delta^{97}Mo$ (‰) | $\delta^{98}Mo$ (‰) | $\delta^{100}Mo$ (‰) | $\delta^{84}Sr$ (‰) | $\delta^{88}Sr$ (‰) | $\delta^{134}Ba$ (‰) | $\delta^{135}Ba$ (‰) | $\delta^{137}Ba$ (‰) | $\delta^{138}Ba$ (‰) |
|---|---|---|---|---|---|---|---|---|---|---|---|---|---|---|---|---|---|---|
| M1-A5-G1112 | AB2 | 1.4×1.5 | 11.1±0.2 | 632±132 | 8±54 | 16±60 | 16±85 | 11±98 | 17±84 | 79±102 | −66±72 | 10±98 | | −32±375 | | | | |
| M1-A9-G353 | AB2 | 1.3×1.2 | 9.6±0.2 | 790±166 | 70±32 | 46±20 | −80±152 | −90±174 | −91±149 | −187±165 | −67±137 | −57±180 | | | | | | |
| M2-A1-G554 | AB2 | 0.7×1.0 | 8.4±0.4 | 1569±188 | 13±34 | 36±44 | −472±192 | −486±226 | −190±253 | −285±275 | −234±220 | −418±254 | | −817±226 | | | | |
| M2-A1-G773 | AB2 | 0.7×0.8 | 8.4±0.4 | 1862±386 | 109±24 | 45±26 | −35±220 | 62±269 | 119±239 | −217±218 | 68±212 | −25±253 | | | | | | |
| M2-A1-G807 | AB2 | 1.0×1.1 | 10.3±0.6 | 633±60 | 56±22 | 20±24 | −106±144 | −166±160 | −180±132 | −159±158 | −143±124 | −127±163 | 152±386 | −9±94 | | | | |
| M3-G348 | AB2 | 0.8×1.0 | 8.6±0.2 | 1401±660 | 34±20 | 62±24 | −23±72 | −41±82 | 1±72 | −24±82 | −16±65 | −4±84 | 116±221 | 27±60 | 13±162 | −28±111 | −23±99 | −34±79 |
| M3-G439 | AB2 | 0.4×0.8 | 10.0±0.2 | 1503±158 | −34±16 | −13±20 | −114±169 | −14±210 | −148±160 | −66±201 | −40±160 | −16±210 | 120±429 | −87±105 | | | | |
| M3-G447 | AB2 | 1.5×0.7 | 11.0±0.2 | 1691±276 | −32±16 | 20±20 | −80±59 | −100±68 | −41±60 | −51±69 | −28±55 | −48±70 | | | | | | |
| M3-G756 | AB2 | 0.5×0.3 | 11.0±0.2 | 984±238 | 73±24 | 61±30 | −250±218 | −382±225 | −275±207 | −127±279 | −104±221 | −193±265 | | | | | | |
| M3-G1356 | AB2 | 0.9×0.6 | 9.6±0.2 | 1271±472 | 110±18 | 63±22 | −2±52 | −34±58 | 3±51 | −3±59 | −15±46 | 18±60 | | −226±291 | | | | |
| M3-G1669 | AB2 | 0.8×0.8 | 8.6±0.2 | 1587±612 | 14±16 | 2±20 | −177±179 | −85±223 | −51±194 | −154±212 | −105±170 | 64±247 | | | | | | |
| M3-G1693 | AB2 | 1.0×0.9 | 10.0±0.2 | 1414±148 | 31±16 | 20±22 | 100±111 | 137±130 | 0±101 | 44±121 | 81±99 | 108±128 | | | | | | |
| M1-A6-G538 | MS | 1.0×1.2 | 49±0.6 | 1745±502 | 68±34 | 88±24 | −724±32 | −664±44 | −425±50 | −438±59 | −140±60 | −697±41 | | 68±280 | 255±527 | −741±138 | −584±148 | −534±109 |
| M2-A1-G651 | MS | 0.8×0.9 | 56±3.2 | 564±84 | 74±20 | 61±22 | −532±83 | −519±104 | −356±102 | −379±120 | −161±110 | −623±88 | | 63±228 | | | | |
| M2-A1-G653 | MS | 0.6×0.7 | 71±4.0 | 316±40 | 46±22 | 33±26 | −550±85 | −507±107 | −286±112 | −183±145 | −36±127 | −542±103 | | | | | | |
| M2-A1-G695 | MS | 0.8×0.7 | 64±3.6 | 2110±252 | 33±20 | 37±22 | −756±71 | −747±88 | −373±122 | −426±138 | −281±120 | −754±89 | −918±60 | 57±55 | 70±463 | −872±89 | −390±182 | −209±174 |
| M2-A2-G189 | MS | 0.6×0.8 | 83±4.0 | 1134±310 | 2±22 | 57±28 | −894±17 | −840±26 | −557±37 | −475±50 | −205±50 | −901±21 | | | −520±776 | −602±519 | −618±330 | −520±776 |
| M2-A2-G725 | MS | 0.8×0.6 | 48±1.2 | | 80±26 | 107±34 | −855±23 | −818±32 | −535±44 | −441±60 | −279±54 | −872±27 | | −129±166 | −422±578 | −797±172 | −456±196 | −415±149 |
| M3-G379 | MS | 1.1×1.3 | 71±3.0 | 457±50 | −12±24 | 33±26 | −377±25 | −364±31 | −224±29 | −181±36 | −100±29 | −384±30 | 39±503 | −23±122 | | | −647±308 | −511±281 |
| M3-G956 | MS | 1.2×0.8 | 81±3.6 | 1499±256 | 54±20 | 54±22 | −362±65 | −406±74 | −234±73 | −120±94 | −105±74 | −406±73 | | | | | | |
| M3-G1116 | MS | 1.1×1.2 | 59±2.6 | 1053±152 | 34±20 | 57±20 | −624±38 | −612±47 | −431±48 | −284±67 | −159±58 | −641±44 | | | | | | |
| M3-G1268 | MS | 0.9×0.9 | 94±4.0 | 1641±196 | −1±20 | 33±20 | −649±39 | −644±49 | −389±54 | −322±70 | −128±64 | −634±48 | | | | | | |
| M3-G1287_1 | MS | 2.1×2.0 | 42±1.8 | 778±204 | 67±22 | 80±24 | −873±7 | −849±10 | −551±15 | −436±21 | −201±21 | −863±10 | | | | | | |
| M3-G1375 | MS | 1.3×1.9 | 51±2.2 | 4039±501 | 77±18 | 104±18 | −635±25 | −611±31 | −403±33 | −281±45 | −185±38 | −580±33 | | 104±150 | −17±331 | −756±89 | −511±122 | −390±101 |
| **M2-A2-G266** [a] | **MS** | **1.0×0.9** | **93±2.2** | | **1±26** | **26±32** | **−24±394** | **62±494** | **405±519** | **−17±453** | **21±368** | **105±493** | | **27±357** | **−21±515** | **−619±183** | **−247±277** | **−47±268** |
| **M2-A2-G569** | **MS** | **0.6×0.8** | **72±4.0** | **434±22** | **-9±18** | **2±22** | **−127±163** | **−80±197** | **−45±171** | **−11±204** | **79±172** | **−117±190** | | **109±329** | **48±729** | **−443±280** | **−319±327** | **−255±224** |
| **M3-G1020** | **MS** | **1.4×0.6** | **53±2.4** | **2429±312** | **115±26** | **88±26** | **−27±48** | **−42±55** | **4±48** | **−27±55** | **3±44** | **−8±56** | **−893±166** | **−105±70** | **558±551** | **−645±140** | **−383±165** | **−286±144** |

[a]: Correlated Mo and Ba isotope ratios of the three highlighted grains (Fig. 4) indicate that large amounts of Mo contamination were sampled during the grain analysis.

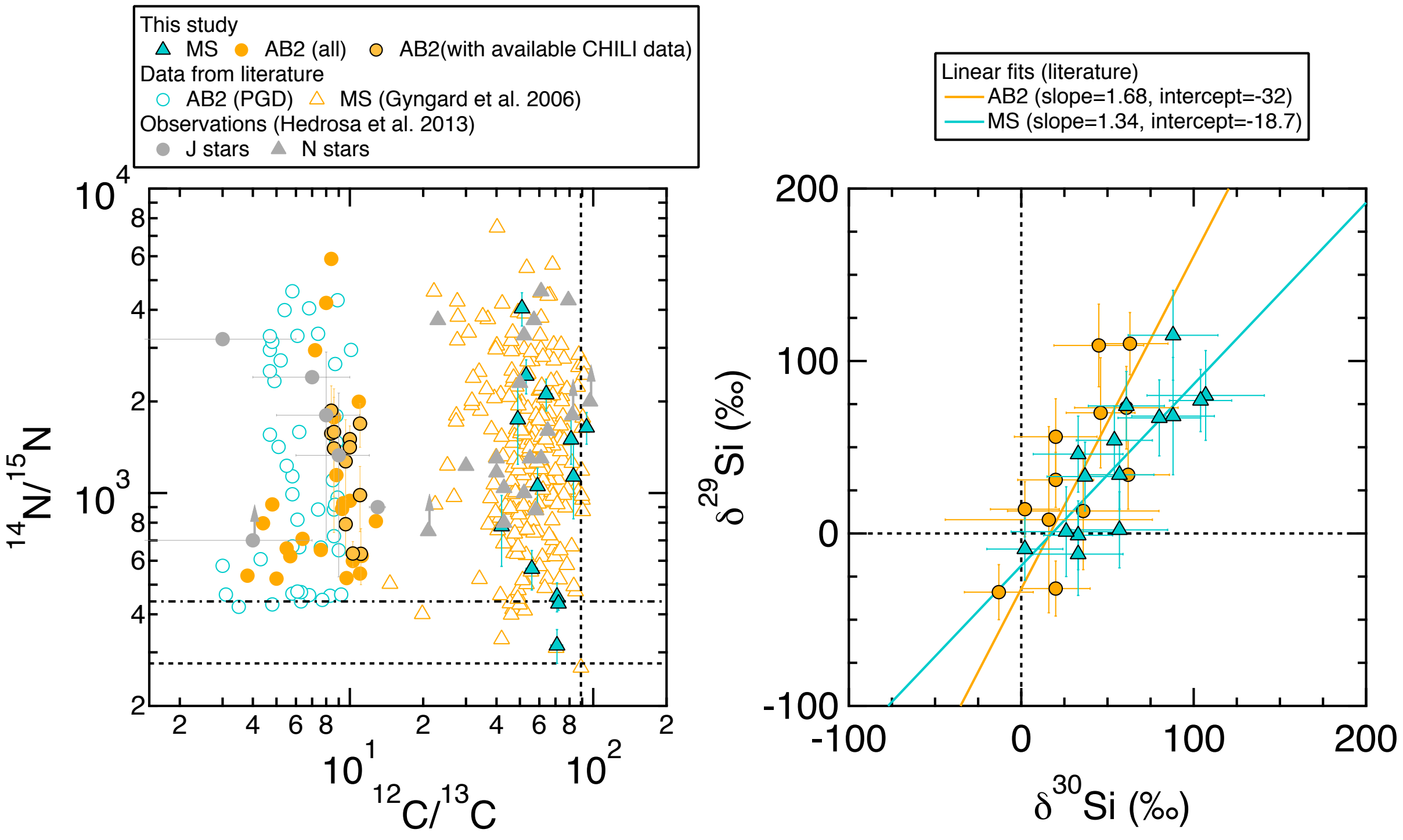

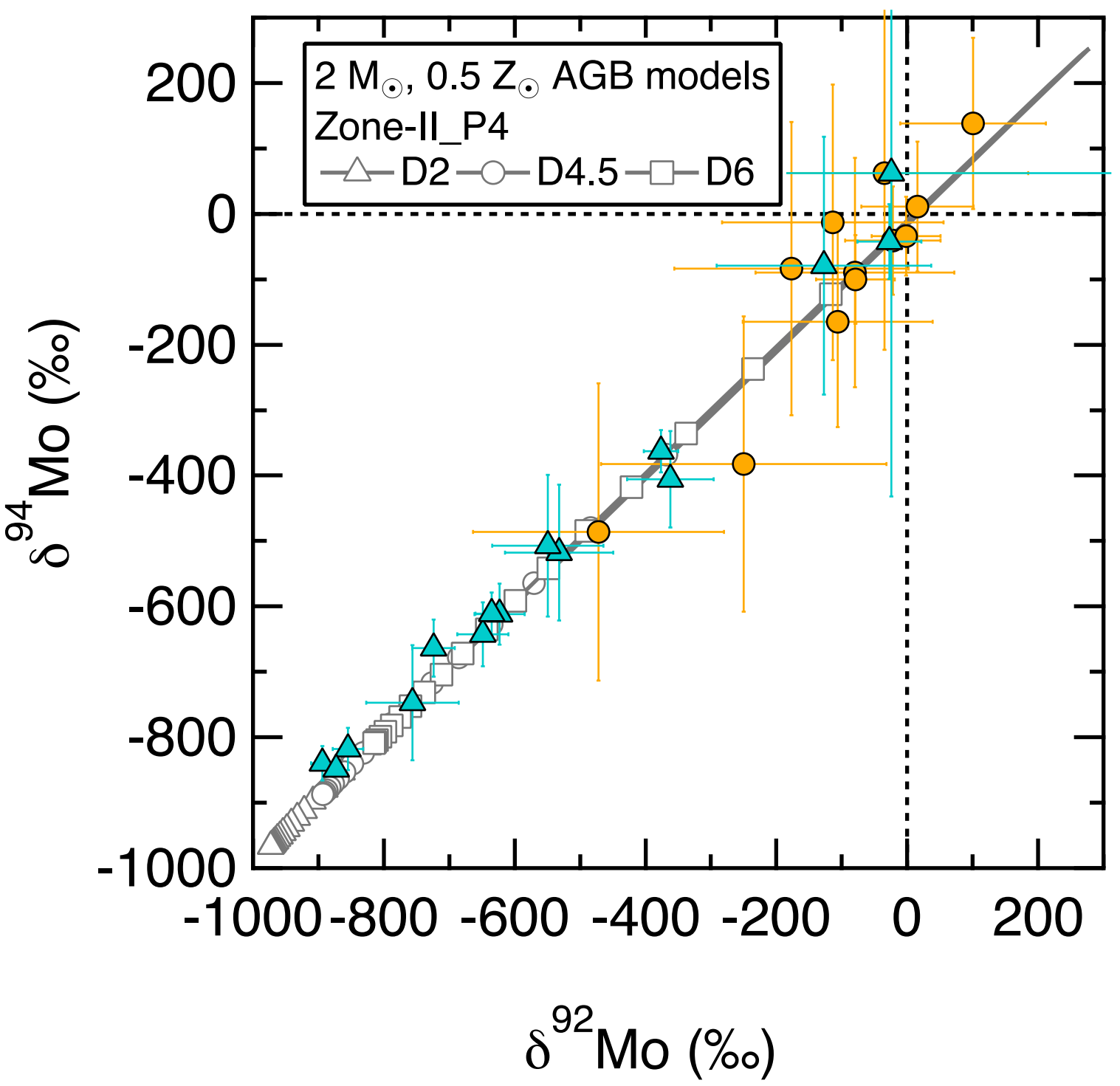

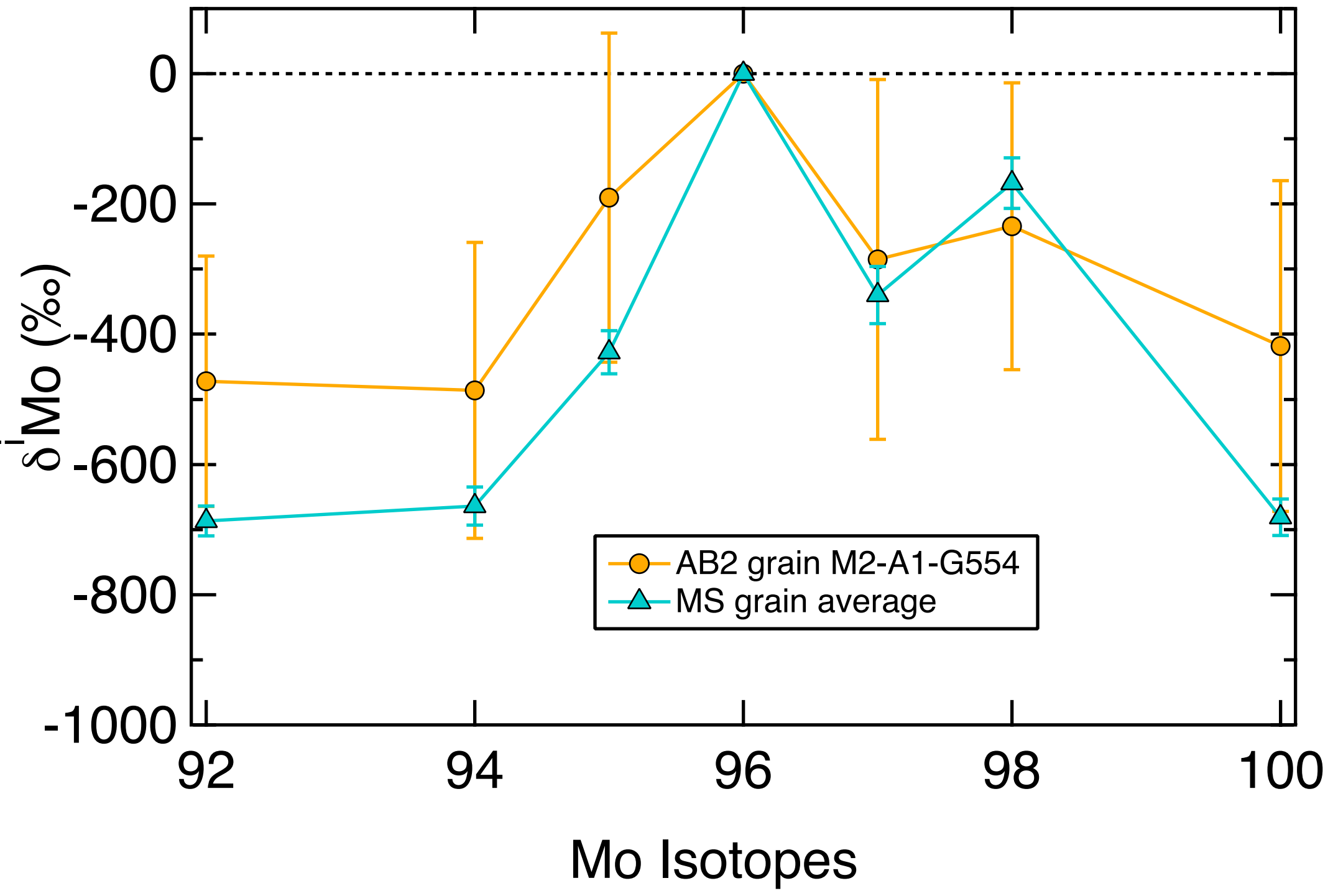

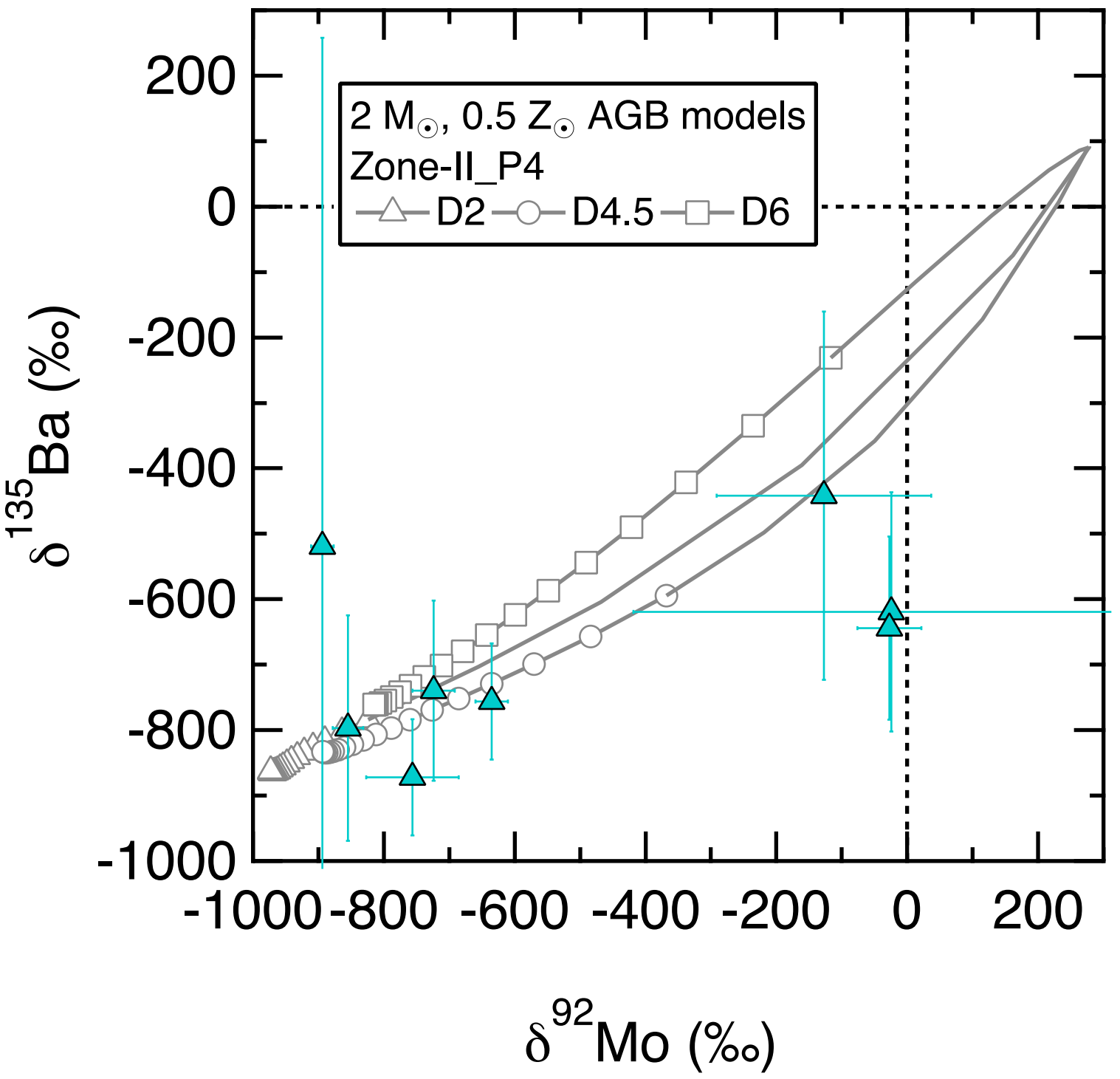